# Direct Mapping Hidden Excited State Interaction Patterns from *ab initio* Dynamics and Its Implications on Force Field Development

Fang Liu, Likai Du*, Dongju Zhang, Jun Gao*

Hubei Key Laboratory of Agricultural Bioinformatics, College of Informatics, Huazhong Agricultural University, Wuhan, 430070, P. R. China

dulikai@mail.hzau.edu.cn; gaojun@sdu.edu.cn

## Abstract

The excited states of polyatomic systems are rather complex, and often exhibit meta-stable dynamical behaviors. Static analysis of reaction pathway often fails to sufficiently characterize excited state motions due to their highly non-equilibrium nature. Here, we proposed a time series guided clustering algorithm to generate most relevant meta-stable patterns directly from *ab initio* dynamic trajectories. Based on the knowledge of these meta-stable patterns, we suggested an interpolation scheme with only a concrete and finite set of known patterns to accurately predict the ground and excited state properties of the entire dynamics trajectories. As illustrated with the example of sinapic acids, the estimation error for both ground and excited state is very close, which indicates one could predict the ground and excited state molecular properties with similar accuracy. These results may provide us some insights to construct an excited state force field with compatible energy terms as traditional ones.

## Introduction

The photophysical or photochemical processes are extremely important for the evolution of life and environments. After the molecule is excited onto higher electronic states, the molecule would undergo a rather complex sequence of dynamics, such as radiative electronic transitions (fluorescence, phosphoresences), nonradiative electronic transitions (internal conversions, intersystem crossings), energy transfers and chemical reactions, etc. Many efforts have been devoted to understand the molecular basis of the possible photophysical or photochemical mechanism in the last decades.[1-10] The reliable theoretical simulation of excited state processes of polyatomic molecules is not so straightforward in most cases, which requires the accurate calculation of electronic excited states for highly non-equilibrium molecular geometries and the nonadiabatic transitions between the electronic and nuclear degrees of freedom[11-15]. The non-equilibrium nature of excited state processes presents some difficulty in establishing a realistic description of these ultrafast processes which are completed in picoseconds or even femtoseconds. Similar problem is also mentioned in the thermally activated ground state reactions, especially, for the transition state structures, the dynamical correlations are known to disrupt the minimal energy path picture.[16-18] Therefore, a dynamic description is much preferred.

*Ab initio* molecular dynamics (AIMD) methods have been extended for the excited states problems[19-24], such as the on-the-fly surface hopping method[21-22, 25-27], for which, the dynamics and electronic structure problems are solved simultaneously. This involves nuclear dynamics to determine the time evolution of the molecular geometry in concert with electronic structure methods capable of computing electronic excited state potential energy surfaces (PES). A large number of trajectories are usually produced from the excited state AIMD simulations. Such simulations directly include all nuclear degrees of freedom, which provide a rather rich picture of the microscopic processes. However, for medium to large size molecules, the trajectories are generally chaotic and becoming inscrutable for human to extract the physical insights of interesting. Therefore, it is very necessary to design more intelligent algorithms to depict not only the available reaction channels, but also further dynamics details, such as the hidden meta-stable states and their interaction networks.

To meet the challenge of tackling with the PES complexity over AIMD trajectories, much recent efforts have been devoted to machine learning (ML) algorithms.[28-35] Generally, the number of local minima, and hence, the number of meta-stable states, grows exponentially along with system size. An important method for shrinking the data set is to apply a clustering algorithm to obtain a family of clusters (microstates) of much smaller size than the original data set. In this aspect, the nonlinear dimensionality reduction algorithms have been used to investigate the conical intersection topology of the excited state dynamics[28]. And the Markov state models (MSMs) have been proposed to automatically construct coarse grained representations for biological macromolecular conformation dynamics[36-37], that are readily humanly understandable. As a practical issue, the prospect of using ML algorithms to tackle the flood of dynamics data to yield statistical significance is indeed very promising.

In recent years, ML algorithms also become a popular and effective tool to improve computational chemistry methods.[38-43] Using ML algorithms to reproduce *ab initio* calculation results would greatly reduce the computational cost without loss of the accuracy[30]. And the ML algorithms have been successfully used to predict various molecular properties on their electronic ground or excited states.[30, 44-45] For example, the ML algorithms have been implemented to predict (PES) at QM and QM/MM level successfully using neural networks (NN) model.[39, 46-47] The ML algorithms are also reported to accelerate the AIMD simulation of material systems.[48-49] In this regard, it is very interesting to construct new efficient computational methods based on the knowledge of the statistical significance from ML studies.

So for, excited state molecular dynamics (MD) are often restricted within direct *ab initio* methods for small to medium size molecules, and the use of parameter-based empirical force field is generally avoided. In contrast, the classical MD with empirical force field has been successfully applied to very large molecule systems in the ground state, such as protein conformational dynamics or protein-ligand interactions.[50-57] A few attempts have been devoted to develop excited state force field for efficiently describing electronic excited states motions. The force fields parameter sets has been developed for a few typical molecules in low-lying electronic excited states based on quantum chemical calculations[58-59]. A few novel model for excited state empirical force field have also been proposed, such as the interpolated mechanics-molecular mechanics (IM/MM)[60], and electron force field (eFF)[61]. And these progresses may provide more insights for much longer time scale (i.e. nanoseconds) excited state simulation of large molecules in condense phases. The main difficulty in developing excited state empirical force fields is the relative scarcity of the universal mathematical function forms (PESs and their couplings) and the failure to sufficiently characterize excited state motions due to their highly non-equilibrium nature. The excited state AIMD trajectories contain large amount of information about their traveling PES, which can be used intrinsically as data sets to design the new models. We suspect the ML algorithm may provide an idea tool for revealing the coarse-grained representations of the excited state processes, as well as the main transitions between the hidden meta-stable states. And the data mining of the AIMD trajectories may promote the future excited state force field development.

In this work, we present a time series guided clustering algorithm to extract the main features of meta-stable states and their correlations from an ensemble of excited state AIMD trajectories. On the basis of these finite meta-stable patterns, the conformation similarity was explored to build an interpolation scheme, namely, the prediction with ensemble models (PEM), to estimate the ground and excited state properties of the entire dynamics trajectories. The PEM method does not require any training data beyond the clustering algorithm, and we could correctly predict the charge population and excitation energy, in comparable with the DFT/TDDFT calculations. As a test case, the excited states $S_1$ of sinapic acid (SA) was used as a benchmark system, which is an essential UV-B screening ingredient in natural plants.[62] This work highlights the potential power of ML algorithm in computational chemistry to extract chemical insights or develop the state-of-art theoretical models.

## Models and Methods

### Dynamics Simulation and Data Sets

The molecular conformation data sets were mainly collected from our previous excited state AIMD simulations of sinapic acids (SA).[62] For simplicity, the *cis*-SA molecule in gas phase is used as a benchmark system (Scheme 1), and we only focus on a single potential surface ($S_1$). This is reasonable since the nonadiabatic decay to the ground state is not observed within the simulation time scale, which is very different from the dynamic behavior of the solvented SA molecule.[62-65] Since the excited state dynamics would stay on a single excited state surface ($S_1$), we can view it as an excited state Born-Oppenheimer molecular dynamics (BOMD) simulation. However, it should be noted that our following protocols can be easily applied to two or more coupled PES conditions.

The molecule spends most of the trajectory time dwelling in a free energy minimum, "waiting" for thermal fluctuations to push the system over a free energy barrier. Thus, it is extremely difficult to adequately sample the conformation space for complex molecules due to the limited timescales accessible for excited state AIMD simulations. Thus, an ensemble of uncoupled AIMD trajectories was used for our subsequent analysis. Totally, 200,000 snapshots from dynamics trajectories were obtained. To improve the sampling efficiency, we only sample the snapshots of local minima along the each trajectory (13226 structures).

The ground state calculations were performed at B3LYP/6-31G(d,p) level, while, the excited state calculations were performed at TD-B3LYP/6-31G(d,p) level. The ground and excited state molecular electrostatic potential (ESP) is also calculated for subsequent analysis. The adopted DFT/TDDFT methods have been carefully calibrated with known experimental evidence in our previous work[62], and can be used for thousands to millions energy and gradient calculations. Dynamics treatment with more accurate electronic-structure and advanced dynamical methods at all atoms level should represent the great challenge for future theoretical chemistry. All electronic structure calculations were performed with Gaussian 09 package[66]. The data sets and some scripts can be obtained upon request or downloaded from https://github.com/dulikai/bidiu.

**The Clustering Algorithms**

The dynamics of complex systems typically involves various meta-stable intermediates. It is necessary to decompose conformation space into a set of kinetically meta-stable states. This can be achieved by various clustering algorithms, such as K-means clustering, mean shift, hierarchical clustering, artificial neural network and etc. Here, the K-means, as a simple and robust algorithm, is used to classify the conformation space into a set of discrete states or clusters, which should be corresponding to basins of attraction of local minima on the PES.

The K-means clustering algorithm aims to partition $n$ observations into $k$ clusters, in which each observation belongs to the cluster with the nearest mean, serving as a prototype of the cluster. The most critical parameter in the K-means clustering is the number ($k$) of the clusters or the centroids. Thus, we have tested many cluster numbers (i.e. k = 4~400). The subsequent analysis verified that 12 clusters are sufficient and efficient. The molecular internal coordinates are used as molecular descriptors for clustering, especially, a few critical dihedral angles[62, 67] are chosen to characterize the excited state motion of the SA molecule (Scheme 1). The protocol is implemented with scikit-learn module[68] in Python.

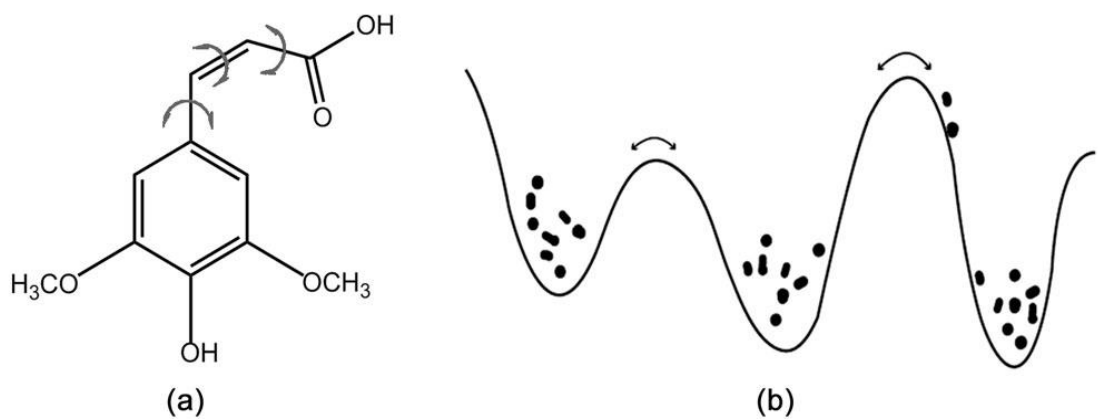

**Scheme 1.** (a) The schematic structure of *cis*-SA molecule and the important dihedral angles are labeled with arrows. (b) Possible distribution of the conformational states along a simple potential energy curve. The points represent the accessible conformational states during dynamics

**The Prediction with Ensemble Models**

The clustering algorithm ensures that similar conformations are grouped into the same meta-stable states or clusters, in a methodical and unbiased fashion. Thus, the dynamics process could be represented by only a limited number of meta-stable patterns. The conformations in the same cluster should have much similar properties (low bias) than among different clusters (high variance). And we can estimated the transition probabilities between meta-stable states by counting the number of transitions along the time series of the trajectories, and thus form a kinetic meta-stable states network, namely time series fusion (TSF) network.

Based on the concept of ensemble averaging[69-71], we suggest an interpolation scheme, namely Prediction with Ensemble Models (PEM) algorithm, to build a model and predict reliable molecular properties, such as charge population and excitation energy. In the machine learning realm, ensemble averaging is one of the simplest types of committee machines, which provides an ideal technique to combine multiple models or patterns to produce a desired output. Usually, an ensemble of models performs better than any individual model, because the various errors of the models can be averaged out[69]. In summary, ensemble averaging creates a group of networks, each with low bias and high variance, and combines them to a new network with low bias and low variance. It is thus a possible solution of the bias-variance dilemma.[69, 71]

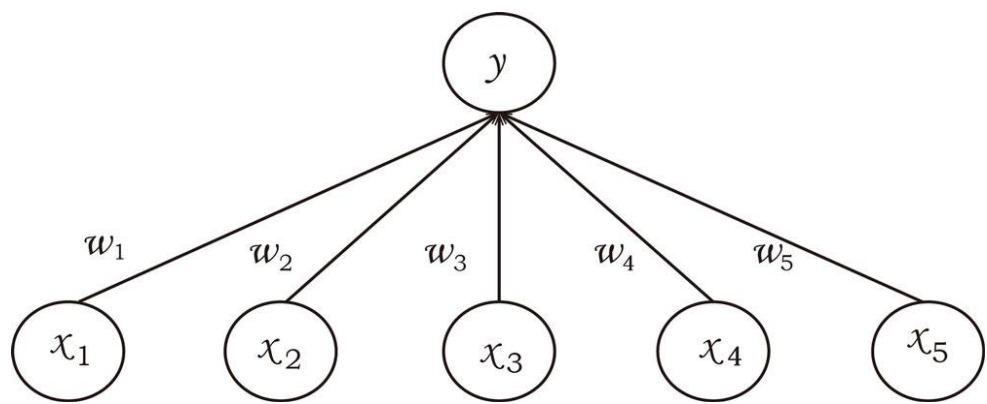

**Scheme 2**. The schematic representation of the ensemble averaging, which can be viewed as an interpolation algorithm.

The theory of ensemble averaging relies on two properties of artificial neural networks: 1) In any network, the bias can be reduced at the cost of increased variance. 2) In a group of networks, the variance can be reduced at no cost to bias. In this work, a general version of ensemble averaging is defined as a weighted sum of finite number of clusters, with low bias and high variance. If the molecule descriptors (**X**) are specified, the estimated result **V** can be defined as

$$V(\mathbf{X}) = \sum_{i=1}^{M} \sum_{j=1}^{n} \omega_{ij}(\mathbf{X}) \mathbf{T}_{ij}(\mathbf{X}) \tag{1}$$

In the above equation, $\mathbf{T}_{ij}$ is the property of one element ($j$) in each cluster ($i$), and the distinct clusters are generated by the K-means algorithm from the time series of trajectories. And M is the number of clusters, $n$ is the number of possible elements in each cluster, and ω is a set of weights.

Here, we use the kernel function in kernel-based ML methods[29] to directly obtain the weights (ω), which tends to be somewhat easier to set up in practice than the artificial neural networks. The kernel function should have the following features: 1) The formula should be continuous in the input space of molecular descriptors, so any small perturbation of the system does not change the results too much. 2) The **X** far from a specific cluster should have less weight, because such cluster has little similarity on the conformation space.

Thus, the weight/kernel function is defined as a function of general distances between an arbitrary geometry and a set of known geometries.

$$\omega_{ij} = \frac{1/u_{ij}}{\sum_{i=1}^{N}(1/u_{ij})^4} \qquad (2)$$

The general distances is defined as

$$u_{ij}(\mathbf{X}) = \sqrt{\sum_{k=1}^{p}(\mathbf{X}-\mathbf{X}_k')^2} \qquad (3)$$

In Eq. (3), $p$ is the number of molecular descriptors (internal coordinates), and $\mathbf{X}_k'$ is known values of molecular descriptors. The above equation is very similar with well-known PES interpolation algorithm, i.e. the Shepard interpolation[72-73]. This is not surprising because much of ML algorithms are just the interpolation between data points, at its core. It should note that the optimization problem of finding the weight ω can also be solved through the training of neural networks, if one does not care about the drawback of their interpretability as a data exploration tool.[74-75]

## Results

### Characters of Possible Excited State Meta-stable Patterns

Here, we focus our attentions on directly extracting the physical insights from excited state dynamics trajectories. To build such kinetic models, it is necessary to map out the dominant long lived, kinetically meta-stable states visited by the molecular system. Thus, we use a conformation clustering algorithm (K-means) to automatically split the time series of dynamics trajectories into geometrically distinct clusters. This also allows us to characterize possible rare events not easily observable in simulations.

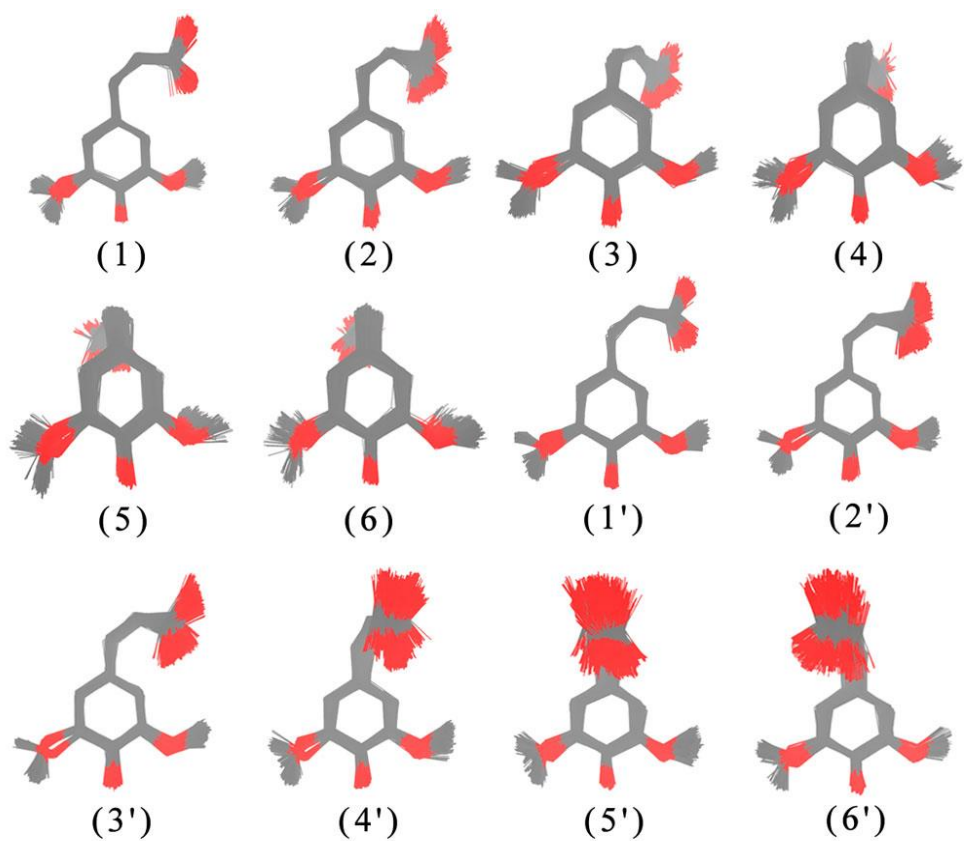

**Figure 1.** Possible meta-stable patterns derived from K-means clustering results. The oxygen and carbon atoms are shown in red and gray, and the position of the hydrogen atom is not shown for clarity.

Figure 1 shows the overlapped molecular geometries for each pattern. The geometric features of meta-stable patterns were sampled on the basis of K-means clustering algorithm. Some attempts have been performed to vary the number (k) of distinct clusters, and finally, the number of k=12 are adopted. It is interesting that the conformation space from dynamics trajectories could be split into a limited number of concrete clusters. The standard deviations of heavy atom coordinates for each pattern after alignments are usually less than 0.5 Å. Because the nearly symmetric character of the dihedral angle motions (Scheme 1a), the clustering results show very similar symmetric features (Figure 1). Two meta-stable patterns (1/1') are mainly in the same plane of the aromatic ring, while some meta-stable patterns (6/6') are perpendicular to the plane of the aromatic ring. Most other patterns show transition characters between these two kind of meta-stable patterns. Since only a few dihedral angles were selected as molecule descriptors, the geometric deviation from the methoxy group is observed. This shows neglectable effects for our qualitative analysis of the excited state *cis*-SA molecule, and thus the clustering results are still very reasonable.

Next, the transition probabilities between these meta-stable patterns are determined, in order to create a kinetic model of the system's conformational dynamics. This can be easily achieved by counting the number of transition times along the time series of the dynamics trajectories. After the transition probabilities between clusters are resolved, one can have a kinetic network, which retains a coarse version of the dynamics. This conformation space network can be drawn as a graph, which can be described by graph theory. Each pattern is represented by a vertex as a colored circle in the graph. And the edge represents the connection among patterns, which could undergo direct transformation between each other.

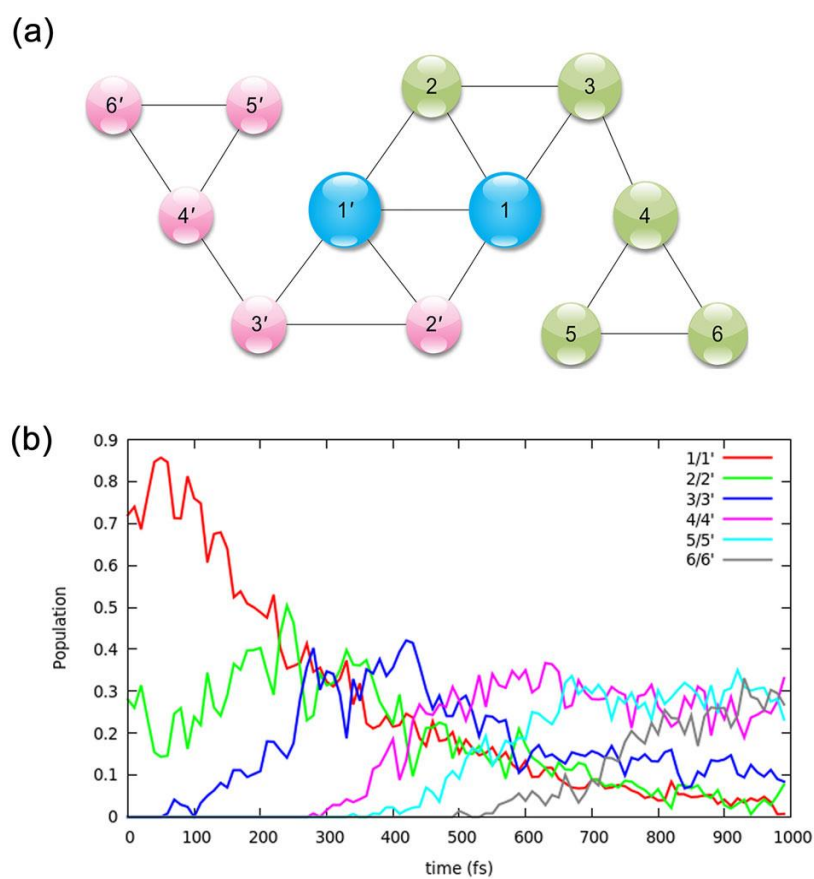

**Figure 2**. (a) The network formed by connecting the meat-stable patterns. The size of the circles reflects the relative number of snapshots in each pattern. (b) The time dependent population of each meta-stable pattern. And the meta-stable patterns could be corresponding to the possible free energy basins.

At first glance, the network is not fully connected (Figure 2a), which indicates there exists explicit reaction pathway among these meta-stable patterns. The near symmetric rotation character of *cis*-SA molecule is also reflected in the network. The network seems to be divided into three sub-graphs; one is correlated to near in-plane patterns, while the other two are correlated to out of plane patterns (i.e. inward and outward of the aromatic ring plane). The patterns 3/3' and 4/4' are vertex separators, the removal of which would disconnect the remaining network. Therefore, this graph provides a suitable view to directly interpret the excited state trajectories, and an acceptable way to recognize the non-negligible intermediate states. We also try to provide a time dependent description of this kinetic model. Figure 2b shows the population of each meta-stable pattern as a function of time. Note that the pattern $i$ and $i'$ (1~6) are combined to provide a simplified representation. Generally, the features of in-plane patterns (1/1' or 2/2') are often found at the beginning of the trajectories, and features of out of plane patterns (i.e. 4/4', 5/5', 6/6') are mainly observed during the evolution of the dynamic trajectories. The initial quick drop of the population of the patterns 1/1' is observed within the first 300 fs. And then, the continuous decay from 20% to 5% takes place along with fast oscillations. The total population of the out of plane patterns (i.e. 4/4', 5/5', 6/6') increases to nearly 80% beyond 600 fs. In this way, the dynamics process could be described using merely a few meta-stable patterns.

Note that this kinetic network is built from short timescale (~1 ps) dynamics trajectories and an ensemble of uncoupled trajectories were used to extract long timescale dynamics features. This ensemble dynamics approach has been commonly used in the classical MD simulations[76-79]. We have also performed five longer timescale (~10 ps) excited state AIMD simulations and these meta-stable patterns are indeed observed in longer excited state AIMD trajectories. Therefore, our algorithm can successfully tackle the issue of capturing proper meta-stable patterns that faithfully represent excited state dynamics at the timescale of interest.

**The Statistical Analysis of the Meta-stable Patterns**

The clustering algorithm creates a group of networks, which gives us a unique perspective for understanding excited states processes. Generally, each cluster shows low structural bias, while different clusters show high structural variance. This means that each meta-stable pattern holds very similar molecular structures. Frequently, the molecular properties of similar structure are very close. Here, we focus our attention on the distribution of molecular properties for each meta-stable pattern, such as excitation energy and charge population, which is also distinguishable among meta-stable patterns.

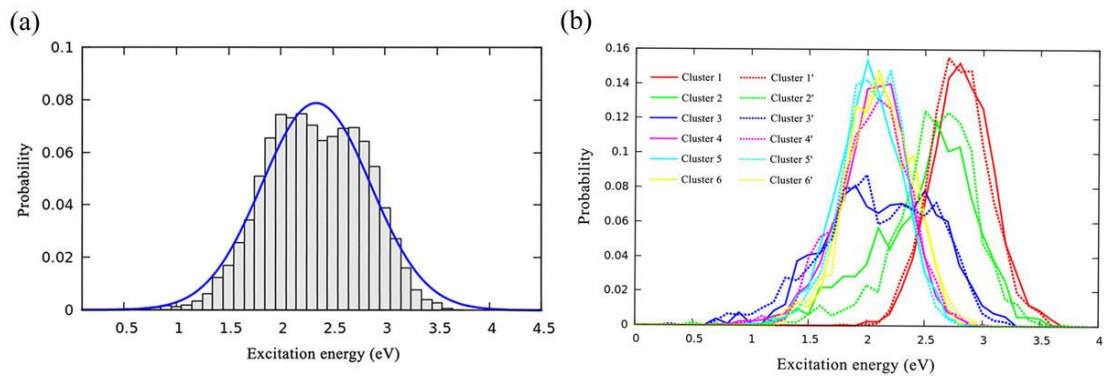

**Figure 3.** (a) The distribution of excited energies for all meta-stable patterns. The fitted Gaussian distribution curve is shown (μ=2.34, σ=0.52). (b) The Gaussian distribution of excited energies for each meta-stable pattern.

The excitation energies are very important to understand the molecular excited states. Figure 3a shows distribution of excitation energies ($S_1$) for all 13226 sampled structures. Significant variation of the excitation energies is observed. The distribution curve is fitted using Gaussian function with the mean value (μ) of 2.34 eV, and standard derivation (σ) of 0.52 eV. Furthermore, the distribution of the excitation energies is also analyzed for each meta-stable pattern (Figure 3b). The mean value and standard derivation of the Gaussian distribution were fitted as indicators. The excitation energies of the patterns 1/1' show the highest mean value of excitation energy (μ=2.8 eV). As the carboxyl moiety rotates out of the aromatic ring plane, the value of excitation energy obviously reduces. The patterns 4/4', 5/5', 6/6' show lower excitation energy (μ=1.9~2.1 eV). The transition patterns (3/3', 4/4') show boarder excitation energy distribution (σ=0.30~0.50 eV), which may cover the characters of the in-plane and out-of-plane geometries. In summary, each pattern has its distinct characters and property. The standard derivation (σ) for each pattern is much smaller (lower bias) than the whole meta-stable patterns network.

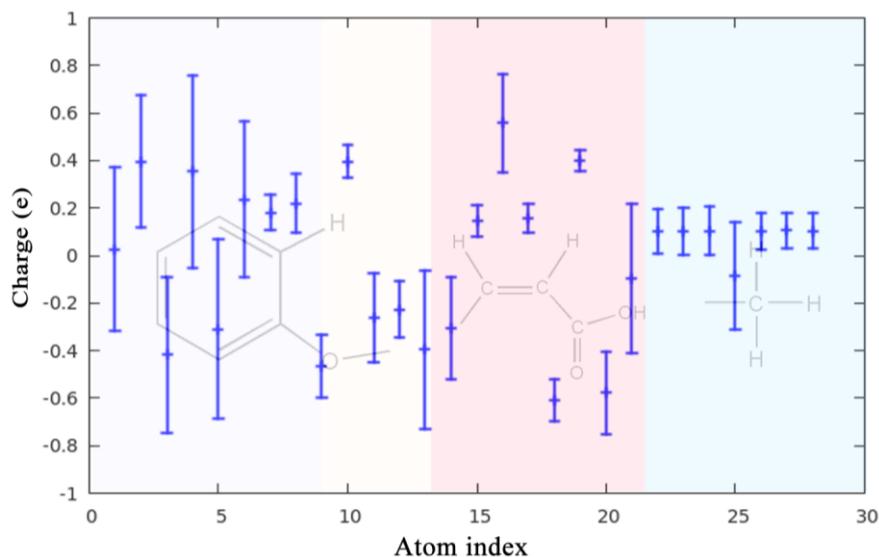

**Figure 4.** The distribution of partial charge for each atom of SA molecule in the meta-stable patterns network. The maximum and minimum of the error bars are related to the value of $\mu \pm 2.58\sigma$ in Gaussian distribution.

Similar conclusion is also available for the ESP partial charge population on the excited state. The partial charge distribution for each atom of SA molecule in 13226 structures has been fitted with Gaussian function. Figure 4a shows the distribution features of the partial charge population for each atom of SA molecule. The median of the error bar refers to the mean value of the Gaussian distribution, meanwhile, the maximum and minimum of the error bar refers to $\mu \pm 2.58\sigma$, which covers 99% of the distribution probability. The partial charge fluctuation for the hydrogen atom is usually neglectable; however, the partial charge for the aromatic ring or the carboxyl motif varies much large. For instance, the partial charge fluctuation is indeed very large, nearly ~1.0 e, for a few atoms. Therefore, it is very necessary to consider the geometric dependence of the partial charge population on the excited state.

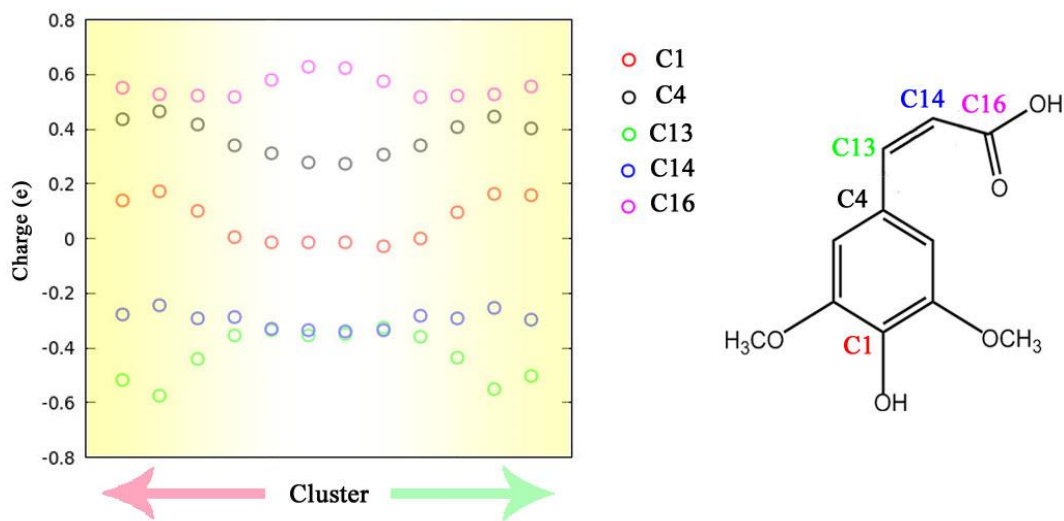

**Figure 5.** The average partial charges of each meta-stable pattern for a few selected atoms of SA molecule. The value of the meta-stable patterns 1/1' are shown in the middle of the X axis.

Then, we also summarized the partial charge population for each meta-stable pattern. For simplify, Figure 5b shows the mean value of partial charges of a few specific atoms with larger variations for each meta-stable pattern. It is obvious that each meta-stable pattern shows distinct partial charge population. The distribution features of the partial charges for each pattern, i.e. mean value ($\mu$) and standard derivation ($\sigma$) are much smaller than that of the entire meta-stable patterns. And each meta-stable pattern shows very different partial charge distribution. In general, the clustering algorithm indeed creates a group of networks, each with low bias and high variance.

**Prediction with Ensemble Models**

In the language of machine learning, the clustering algorithm serves as a classifier to partition the molecular conformations into a few distinct meta-stable patterns. The PEM algorithm requires the number ($M$) of meta-stable patterns for prediction (see Eq. 1). The value of $M$ used for prediction could be equal to the total number of meta-stable patterns, since there is only a finite set of distinct patterns (i.e. $M=12$) for SA molecule. If the number of distinct patterns is very large (i.e. a few thousands), we can optionally reduce this number by screening the value of the general distance, which is inversely proportional to the contribution of each pattern to the molecular properties.

The number ($n$) of possible samples in each meta-stable pattern is another critical parameter to be determined in the PEM algorithm. Two possible solutions can be considered.

1) All elements in each pattern are used, namely, "batch" option, for which all samples are used for predicting molecular properties. In this case, the number $n$ may be very large, i.e. a few thousands for SA molecule. This would strongly lower the efficiency of the PEM algorithm.

2) Only one element in each meta-stable pattern is used, namely, "stochastic" option, for which randomly select only one sample from each pattern, or use the average structure of each pattern. In this case, the number $n$ should be equal to 1. This option is much faster, however, the convergence to the optimal value is too oscillation and stochastic.

In order to overcome the defects of both options, we find a trade-off between the efficiency and reliable, for which a small set of the samples (i.e. $n$=10) in each pattern is used for PEM algorithm, namely "mini-batch" option. This option could reduce the total number of numerical calculations, and also reduce the stochastic behavior. Similar ideas have been commonly used in the realm of machine learning.[43, 80] The PEM algorithm typically scales as O (M·$n$), whereas M and $n$ are very small constant value in the "mini-batch" option. Thus, this algorithm should be significantly faster than electronic structure approaches, such as TDDFT.

It is also very important to determine the uncertainty in the prediction, so one can evaluate the confidence of the results. In the following calculations, the conformations of validation sets (1000 geometries) were randomly sampled from dynamics trajectories. The "mini-batch" option is used with M=12 and $n$=10. This means that all of meta-stable patterns (12 clusters) were used to estimate the molecule properties of any unknown geometry, meanwhile, we randomly select ten geometries ($n$=10) from each pattern to build parameter sets. Note that no training sets were required for the PEM algorithm beyond the clustering algorithm.

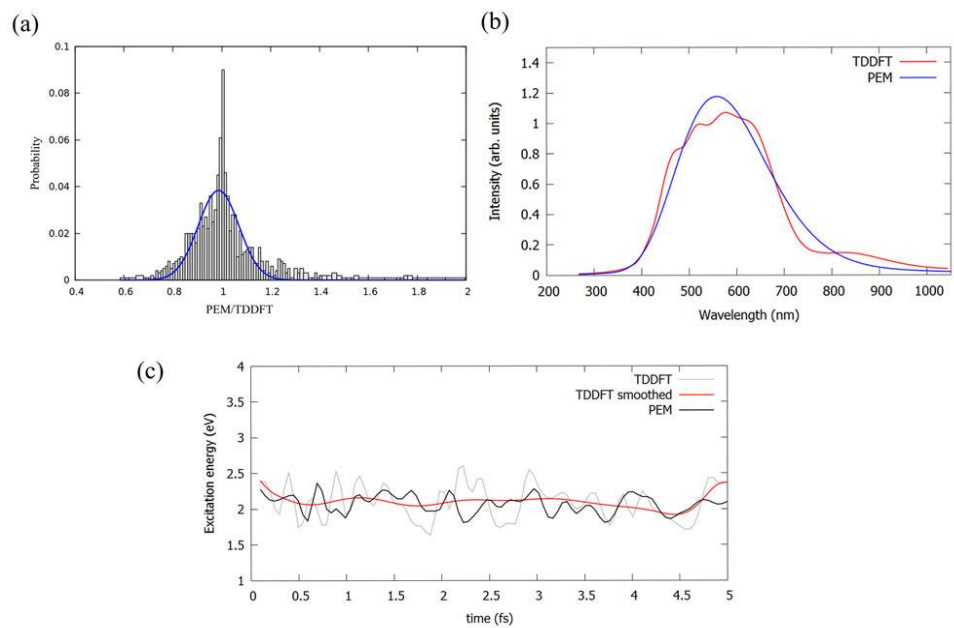

**Figure 6.** (a) The ratio distribution of the excited energy (ratio = PEM *v.s.* TDDFT). The blue curve is fitted with Gaussian function (μ=0.99, σ=0.08). (b) The fluorescence emission spectra roughly estimated from PEM and TDDFT. (c) The time evolution of the excitation energy calculated at PEM and TDDFT level. The red line is the TDDFT result smoothed with Bézier curve, in order to remove the fast oscillations.

Figure 6a shows the performance of excitation energy prediction for $S_1$ state. The distribution of the ratio between the PEM and TDDFT excitation energies is given. The distribution curve is fitted to a Gaussian distribution (μ=0.99, σ=0.08), although the distribution is even more sharp than a Gaussian distribution. The derivation of PEM excitation energy from the TDDFT is within 0.1~0.2 eV (99% cases), which is acceptable even for electronic structure calculations. More sophisticated designing of molecular descriptors or clustering algorithm may improve the prediction. Then, the PEM algorithm is used to estimate the fluorescence emission spectroscopy. In order to take into account of dynamical effects (*vibronic*), the emission spectroscopy is calculated with 100 snapshots sampled from the last 5 ps of five excited state AIMD trajectories. Note also that in all the cases fluorescence was considered to happen only from the first excited singlet state following the so called Kasha's rule[81]. Figure 6b shows the emission spectroscopy obtained from the PEM and TDDFT. The band shape of the emission spectroscopy at PEM and TDDFT level is very close.

Then, the PEM algorithm is used to predict the excitation energies along with a specific excited state AIMD trajectory, which are not used to construct the meta-stable patterns. Figure 6c shows the time evolution for $S_1$ excitation energy from the PEM and TDDFT results for a ~5 ps dynamics trajectory. It seems that the rough trends of time dependent excited energy can be reasonably predicted by the PEM method, however, the details of the excitation energy fails to be predicted. This is reasonable, because our PEM algorithm only includes a few dihedral angles as molecule descriptors. So, the contribution from the fast degree of freedoms is averaged out in this model. More consistent results can be anticipated if more sophisticated molecular descriptors are adopted. In general, excitation energy could be properly described by an ensemble model with a finite set of molecular conformations.

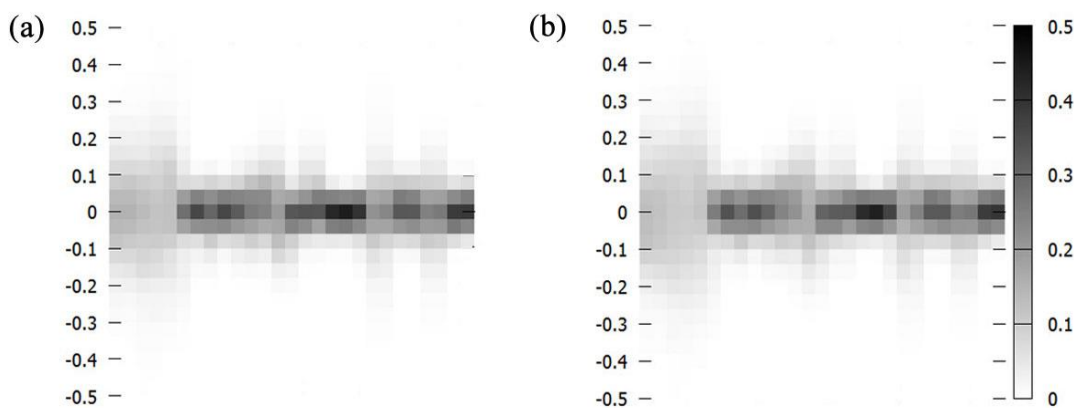

**Figure 7.** Partial charge deviation between PEM and DFT/TDDFT calculations for ground state (a) and excited state (b), respectively. X axis is the atom index for SA molecule.

Figure 7 shows the distribution of the partial charge deviation between the PEM and DFT/TDDFT calculations for the ground state and the excited state (S1). At first glance, the PEM partial charges are improved significantly, and the deviation from the DFT/TDDFT is often less than 0.05 e, in contrast to ~0.50 e in Figure 4. The symmetric fluctuation of the charge deviation around the zero value also suggests that the error cancellation effects may enhance the robust of the PEM partial charges at long timescale dynamics simulation. Another interesting result of the PEM algorithm is that the excited state partial charge is predicted with similar accuracy as the ground state partial charge, considering the very similar distribution of the partial charge deviation in Figure 7. This superiority allow us to update the ground and excited state partial charge at the same foot, although the electronic partial charge of excited state is more complex in comparison with ground state.

Beyond the symmetric fluctuation of the partial charge deviation, the strength of the charge fluctuation for each atom is not the same, for example, the description of the aromatic ring is not very good. This is expected, since the molecular descriptors mainly focus on a few dihedral angles, which only provide a rough representation of the slow degree of freedoms. Therefore, more molecular descriptors may be required to take into account the fast degree of freedoms in the future clustering algorithm. In contrast to the fixed point charge model, the PEM point charge provide a much flexible description of the ground and excited state charge fluctuation (Figure 4 and 7). Although the electronic structure calculations is performed at DFT/TDDFT level in this work, the PEM method can be also applied to any level of electronic structure methods. The use of the PEM method offers a significant improvement in the quality and applicability of electrostatically determined partial charges. In summary, the PEM method provides an alternative choice to directly incorporate the polarization or charge transfer effects on predicting partial charge population.

**Discussions**

The PEM method includes a classifier to identify the possible meta-stable patterns as molecular features, and a predictor to construct the molecular properties for any unknown molecule structure. We have turned out that model ensembles are usually more accurate than any single model, and they are typically more fault tolerant than single models. Therefore, the performance of PEM can be steadily improved by taking into more reference data or optimizing the kernel function. And more sophisticated and intelligent algorithms should also be helpful, i.e. replace the K-means clustering classifier by an auto-encoder or even an artificial neutral network. However, such treatment would scarify the clear physical meaning of variables in this work, and thus the dedicated balance between the performance and human interpretability is of paramount interest.[43]

The simple analytical form of the PEM algorithm provides an alternative choice or procedure to take into account of the large variation of the excitation state molecular properties, i.e. charge population and excitation energy. The good performance of the PEM method may give us some inspiration on the empirical force field development. Theoretically, the description of both molecular ground and excited states should be possible with empirical force fields, especially, when the potential is intended for an application toward adiabatic dynamics on a single surface. However, the "parameters" in the empirical potentials can be vastly different along the molecular conformation for the excited state. And more sophisticated empirical force field model or complex analytic functions may be required.

We suggest that the "parameters" in empirical force field can be constructed on top of the PEM algorithm. Practically, the force field parameters can be assigned for each kind of meta-stable patterns. For instance, the widely used point charge model could be applied for both ground state and excited state, and the real dependencies of partial point charges on molecular conformation may be handled by combining several or even thousands of meta-stable patterns into a single, new point charge predicting model (Eq. 1). Therefore, the requirement to incorporate polarization or even charge transfer into the standard pair-wise potentials can be easily achieved in PEM algorithm, which is compatible with traditional force field. There are two practical advantages. 1) There is no need to construct a complex analytic model; this procedure is cumbersome and computational inefficiency for large molecules. 2) With use of the PEM algorithm, the computer program is very similar with the available empirical force field. This model is fast enough to allow millions of calculations along the dynamics propagation with adequate accuracy. The performance of this assumption would be reported in our continuing work.

The main disadvantage of this approach is the requirements of fully searching the most possible meta-stable patterns. The good news is that the PEM algorithm could easily incorporate more meta-stable patterns of some specific conformation space in a fast iterated procedure, meanwhile, the performance of another part of conformation space would not be affected if the kernel function is properly defined. So, it is highly extensible and flexible. Thanks to the increasing computing power, modern dynamics simulations can easily generate data sets with millions of configurations from an ensemble of uncoupled dynamics trajectories. We also note that similar machine learning based approaches have also been used to accelerate the AIMD simulation of material systems[48-49], because such approach avoids the repeated *ab initio* calculations for the same molecule with similar conformations.

# Conclusions

Our results strongly suggest that ensemble models together with a proper classifier for model selection provides a useful research tool to gain insights from time series of *ab initio* dynamics, as demonstrated in the excited state studies of the SA molecule. Data mining of dynamics trajectories could gain a direct view of the possible meta-stable patterns and their relationship on the physical or biological processes of interest, with only a few commonsense rules. We further suggest that the "state-of-art" PEM method shows good performance in predicting ground and excited state molecular properties, in comparable to DFT/TDDFT calculations. The PEM could sufficiently characterize the feature of excited state motions, and naturally form knowledge based data sets. And the performance of PEM method could steadily be improved in a fast iterated procedure. The PEM method may provide us an alternative perspective to construct excited state force field with similar function form as the ground state one, without using much advance knowledge of the molecule details in the excited states. Its generality and ease of implementation should make it useful in various situations. Further work is going on to investigate the dynamic dependence of the inter-atomic potential itself and its realistic applications on the molecular dynamics simulation.

# Acknowledgements

The work is supported by National Natural Science Foundation of China (Nos. 21503249, 21373124), and Huazhong Agricultural University Scientific & Technological Self-innovation Foundation (Program No.2015RC008), and Project 2662016QD011 and 2662015PY113 Supported by the Fundamental Founds for the Central Universities. The authors also thank the support of Special Program for Applied Research on Super Computation of the NSFC-Guangdong Joint Fund (the second phase)

# References

1. Stolow, A.; Bragg, A. E.; Neumark, D. M., Femtosecond time-resolved photoelectron spectroscopy. *Chem. Rev.* **2004,** *104* (4), 1719-57.
2. Miller, R. J., Femtosecond crystallography with ultrabright electrons and x-rays: capturing chemistry in action. *Science* **2014,** *343* (6175), 1108-16.
3. Wang, S.; Wang, X., Multifunctional Metal-Organic Frameworks for Photocatalysis. *Small* **2015,** *11* (26), 3097-112.
4. Liu, Y. H.; Lan, S. C.; Zhu, C.; Lin, S. H., Intersystem Crossing Pathway in Quinoline-Pyrazole Isomerism: A Time-Dependent Density Functional Theory Study on Excited-State Intramolecular Proton Transfer. *J. Phys. Chem. A* **2015,** *119* (24), 6269.
5. Linic, S.; Aslam, U.; Boerigter, C.; Morabito, M., Photochemical transformations on plasmonic metal nanoparticles. *Nat. Mater.* **2015,** *14* (6), 567-576.
6. Padalkar, V. S.; Seki, S., Excited-state intramolecular proton-transfer (ESIPT)-inspired solid state emitters. *Chem. Soc. Rev.* **2015,** *45* (1), 169-202.
7. Ruban, A. V., Nonphotochemical Chlorophyll Fluorescence Quenching: Mechanism and Effectiveness in Protecting Plants from Photodamage. *Plant Physiology* **2016,** *170* (4), 1903-1916.
8. Chen, M.; Zhong, M.; Johnson, J. A., Light-controlled radical polymerization: Mechanisms, methods, and applications. *Chem. Rev.* **2016,** *116* (17), 10167-10211.
9. Jara‐Cortés, J.; Guevara‐Vela, J. M.; Martín Pendás, Á.; Hernández‐Trujillo, J., Chemical bonding in excited states: Energy transfer and charge redistribution from a real space perspective. *J. Comput. Chem.* **2017,** *38* (13), 957-970.
10. Zhao, G.-J.; Han, K.-L., Hydrogen Bonding in the Electronic Excited State. *Acc. Chem. Res.* **2012,** *45* (3), 404-413.
11. Pisana, S.; Lazzeri, M.; Casiraghi, C.; Novoselov, K. S.; Geim, A. K.; Ferrari, A. C.; Mauri, F., Breakdown of the adiabatic Born–Oppenheimer approximation in graphene. *Nat. Mater.* **2007,** *6* (3), 198-201.
12. Middleton, C. T.; de La Harpe, K.; Su, C.; Law, Y. K.; Crespo-Hernández, C. E.; Kohler, B., DNA excited-state dynamics: from single bases to the double helix. *Annu. Rev. Phys. Chem.* **2009,** *60*, 217-239.
13. González, L.; Escudero, D.; Serrano‐Andrés, L., Progress and challenges in the calculation of electronic excited states. *ChemPhysChem* **2012,** *13* (1), 28-51.
14. Akimov, A. V.; Neukirch, A. J.; Prezhdo, O. V., Theoretical insights into photoinduced charge transfer and catalysis at oxide interfaces. *Chem. Rev.* **2013,** *113* (6), 4496-4565.
15. Zhang, D. H.; Guo, H., Recent advances in quantum dynamics of bimolecular reactions. *Annu. Rev. Phys. Chem.* **2016,** *67*, 135-158.
16. Ammal, S. C.; Yamataka, H.; Aida, M.; Dupuis, M., Dynamics-driven reaction pathway in an intramolecular rearrangement. *Science* **2003,** *299* (5612), 1555-1557.
17. Lourderaj, U.; Park, K.; Hase, W. L., Classical trajectory simulations of post-transition state dynamics. *Int. Rev. Phys. Chem.* **2008,** *27* (3), 361-403.
18. Sun, L.; Song, K.; Hase, W. L., A S(N)2 reaction that avoids its deep potential energy minimum. *Science* **2002,** *296* (5569), 875-8.
19. Bennun, M., Nonadiabatic molecular dynamics: Validation of the multiple spawning method for a multidimensional problem. *J. Chem. Phys.* **1998,** *108* (17), 7244-7257.

20. Bennun, M.; Jason Quenneville, A.; Martínez, T. J., Ab Initio Multiple Spawning: Photochemistry from First Principles Quantum Molecular Dynamics. *J. Phys. Chem. A* **2000,** *104* (22), 5161-5175.
21. Li, X.; Tully, J. C.; Schlegel, H. B.; Frisch, M. J., Ab initio Ehrenfest dynamics. *J. Chem. Phys.* **2005,** *123* (8), 084106.
22. Bittner, E. R.; Rossky, P. J., Quantum decoherence in mixed quantum‐classical systems: Nonadiabatic processes. *J. Chem. Phys.* **1995,** *103* (18), 8130-8143.
23. Zhu, C.; Jasper, A. W.; Truhlar, D. G., Non-Born–Oppenheimer trajectories with self-consistent decay of mixing. *J. Chem. Phys.* **2004,** *120* (12), 5543-57.
24. Yarkony, D. R., Conical Intersections: Their Description and Consequences. In *Conical Intersections :Theory, Computation and Experiment*, Wolfgang Domcke, D. R. Y., Horst Köppel, Ed. World Scientific: 2004; pp 41-128.
25. Miller, W. H.; George, T. F., Semiclassical Theory of Electronic Transitions in Low Energy Atomic and Molecular Collisions Involving Several Nuclear Degrees of Freedom. *J. Chem. Phys.* **1972,** *56* (11), 5637-5652.
26. Tully, J. C., Molecular dynamics with electronic transitions. *J. Chem. Phys.* **1990,** *93* (2), 1061-1071.
27. Craig, C. F.; Duncan, W. R.; Prezhdo, O. V., Trajectory Surface Hopping in the Time-Dependent Kohn-Sham Approach for Electron-Nuclear Dynamics. *Phys. Rev. Lett.* **2005,** *95* (16), 163001.
28. Virshup, A. M.; Chen, J.; Martínez, T. J., Nonlinear dimensionality reduction for nonadiabatic dynamics: The influence of conical intersection topography on population transfer rates. *J. Chem. Phys.* **2012,** *137* (22), 22A519.
29. Rupp, M., Machine learning for quantum mechanics in a nutshell. *Int. J. Quantum Chem.* **2015,** *115* (16), 1058-1073.
30. Hase, F.; Valleau, S.; Pyzer-Knapp, E.; Aspuru-Guzik, A., Machine learning exciton dynamics. *Chem. Sci.* **2016,** *7* (8), 5139-5147.
31. Sánchez-Lengeling, B.; Aspuru-Guzik, A., Learning More, with Less. *ACS Central Science* **2017,** *3* (4), 275-277.
32. Tenenbaum, J. B.; Silva, V. D.; Langford, J. C. In *A global geometric for nonlinear dimensionality reduction*, 2000; pp 2319-23.
33. Singer, A.; Erban, R.; Kevrekidis, I. G.; Coifman, R. R., Detecting intrinsic slow variables in stochastic dynamical systems by anisotropic diffusion maps. *Proc. Natl. Acad. Sci. U. S. A.* **2009,** *106* (38), 16090-5.
34. Shaw, D. E.; Maragakis, P.; Lindorfflarsen, K.; Piana, S.; Dror, R. O.; Eastwood, M. P.; Bank, J. A.; Jumper, J. M.; Salmon, J. K.; Shan, Y., Atomic-level characterization of the structural dynamics of proteins. *Science* **2010,** *330* (6002), 341-6.
35. Maragakis, P.; van der Vaart, A.; Karplus, M., Gaussian-Mixture Umbrella Sampling. *J. Phys. Chem. B* **2009,** *113* (14), 4664-4673.
36. Sirur, A.; De, S. D.; Best, R. B., Markov state models of protein misfolding. *J. Chem. Phys.* **2016,** *144* (7), 075101.
37. Harrigan, M. P.; Sultan, M. M.; Hernández, C. X.; Husic, B. E.; Eastman, P.; Schwantes, C. R.; Beauchamp, K. A.; McGibbon, R. T.; Pande, V. S., MSMBuilder: Statistical Models for Biomolecular Dynamics. *Biophys. J.* **2017,** *112* (1), 10-15.
38. Snyder, J. C.; Rupp, M.; Hansen, K.; Blooston, L.; Müller, K.-R.; Burke, K., Orbital-free bond breaking via machine learning. *J. Chem. Phys.* **2013,** *139* (22), 224104.

39. Bartók, A. P.; Payne, M. C.; Kondor, R.; Csányi, G., Gaussian approximation potentials: the accuracy of quantum mechanics, without the electrons. *Phys. Rev. Lett.* **2010,** *104* (13), 136403.
40. Rupp, M.; Tkatchenko, A.; Müller, K. R.; von Lilienfeld, O. A., Fast and accurate modeling of molecular atomization energies with machine learning. *Phys. Rev. Lett.* **2012,** *108* (5), 058301.
41. Pozun, Z. D.; Hansen, K.; Sheppard, D.; Rupp, M.; Müller, K.-R.; Henkelman, G., Optimizing transition states via kernel-based machine learning. *J. Chem. Phys.* **2012,** *136* (17), 174101.
42. Handley, C. M.; Popelier, P. L., Potential energy surfaces fitted by artificial neural networks. *J. Phys. Chem. A* **2010,** *114* (10), 3371-83.
43. Goh, G. B.; Hodas, N. O.; Vishnu, A., Deep learning for computational chemistry. *J. Comput. Chem.* **2017,** *38* (16), 1291-1307.
44. Hansen, K.; Montavon, G.; Biegler, F.; Fazli, S.; Rupp, M.; Scheffler, M.; Von Lilienfeld, O. A.; Tkatchenko, A.; Müller, K.-R., Assessment and validation of machine learning methods for predicting molecular atomization energies. *J. Chem. Theory Comput.* **2013,** *9* (8), 3404-3419.
45. Ramakrishnan, R.; Dral, P. O.; Rupp, M.; von Lilienfeld, O. A., Big data meets quantum chemistry approximations: the Δ-machine learning approach. *J. Chem. Theory Comput.* **2015,** *11* (5), 2087-2096.
46. Behler, J., Neural network potential-energy surfaces in chemistry: a tool for large-scale simulations. *Phys. Chem. Chem. Phys.* **2011,** *13* (40), 17930-17955.
47. Shen, L.; Wu, J.; Yang, W., Multiscale Quantum Mechanics/Molecular Mechanics Simulations with Neural Networks. *J. Chem. Theory Comput.* **2016,** *12* (10), 4934-4946.
48. Botu, V.; Ramprasad, R., Adaptive machine learning framework to accelerate ab initio molecular dynamics. *Int. J. Quantum Chem.* **2015,** *115* (16), 1074-1083.
49. Li, Z.; Kermode, J. R.; De Vita, A., Molecular dynamics with on-the-fly machine learning of quantum-mechanical forces. *Phys. Rev. Lett.* **2015,** *114* (9), 096405.
50. Cornell, W. D.; Cieplak, P.; Bayly, C. I.; Gould, I. R.; Merz, K. M.; Ferguson, D. M.; Spellmeyer, D. C.; Fox, T.; Caldwell, J. W.; Kollman, P. A., A second generation force field for the simulation of proteins, nucleic acids, and organic molecules. *J. Am. Chem. Soc.* **1995,** *117* (19), 5179-5197.
51. Van Duin, A. C.; Dasgupta, S.; Lorant, F.; Goddard, W. A., ReaxFF: a reactive force field for hydrocarbons. *J. Phys. Chem. A* **2001,** *105* (41), 9396-9409.
52. Marrink, S. J.; Risselada, H. J.; Yefimov, S.; Tieleman, D. P.; De Vries, A. H., The MARTINI force field: coarse grained model for biomolecular simulations. *J. Phys. Chem. B* **2007,** *111* (27), 7812-7824.
53. Henzler-Wildman, K.; Kern, D., Dynamic personalities of proteins. *Nature* **2007,** *450* (7172), 964-972.
54. Xie, W.; Orozco, M.; Truhlar, D. G.; Gao, J., X-Pol Potential: An Electronic Structure-Based Force Field for Molecular Dynamics Simulation of a Solvated Protein in Water. *J. Chem. Theory Comput.* **2009,** *5* (3), 459-467.
55. Bissantz, C.; Kuhn, B.; Stahl, M., A medicinal chemist's guide to molecular interactions. *J. Med. Chem.* **2010,** *53* (14), 5061-5084.
56. Dill, K. A.; MacCallum, J. L., The protein-folding problem, 50 years on. *Science* **2012,** *338* (6110), 1042-1046.
57. Brooks, B. R.; Bruccoleri, R. E.; Olafson, B. D.; States, D. J.; Swaminathan, S.; Karplus, M., CHARMM: A program for macromolecular energy, minimization, and dynamics calculations. *J. Comput. Chem.* **1983,** *4* (2), 187-217.
58. Song, C. I.; Rhee, Y. M., Development of force field parameters for oxyluciferin on its electronic ground and excited states. *Int. J. Quantum Chem.* **2011,** *111* (15), 4091-4105.

59. Ando, K., Excited state potentials and ligand force field of a blue copper protein plastocyanin. *J. Phys. Chem. B* **2004,** *108* (12), 3940-3946.

60. Park, J. W.; Rhee, Y. M., Interpolated Mechanics–Molecular Mechanics Study of Internal Rotation Dynamics of the Chromophore Unit in Blue Fluorescent Protein and Its Variants. *J. Phys. Chem. B* **2012,** *116* (36), 11137-11147.

61. Su, J. T.; Goddard III, W. A., Excited electron dynamics modeling of warm dense matter. *Phys. Rev. Lett.* **2007,** *99* (18), 185003.

62. Liu, F.; Du, L.; Lan, Z.; Gao, J., Hydrogen bond dynamics governs the effective photoprotection mechanism of plant phenolic sunscreens. *Photochem. Photobiol. Sci.* **2017,** *16* (2), 211-219.

63. Baker, L. A.; Horbury, M. D.; Greenough, S. E.; Allais, F.; Walsh, P. S.; Habershon, S.; Stavros, V. G., Ultrafast photoprotecting sunscreens in natural plants. *J. Phys. Chem. Lett.* **2015,** *7* (1), 56-61.

64. Tan, E. M.; Hilbers, M.; Buma, W. J., Excited-state dynamics of isolated and microsolvated cinnamate-based UV-B sunscreens. *J. Phys. Chem. Lett.* **2014,** *5* (14), 2464-2468.

65. Stavros, V. G., Photochemistry: A bright future for sunscreens. *Nat. Chem.* **2014,** *6* (11), 955-956.

66. Frisch, M. J.; Trucks, G. W.; Schlegel, H. B.; Scuseria, G. E.; Robb, M. A.; Cheeseman, J. R.; Scalmani, G.; Barone, V.; Mennucci, B.; Petersson, G. A.; Nakatsuji, H.; Caricato, M.; Li, X.; Hratchian, H. P.; Izmaylov, A. F.; Bloino, J.; Zheng, G.; Sonnenberg, J. L.; Hada, M.; Ehara, M.; Toyota, K.; Fukuda, R.; Hasegawa, J.; Ishida, M.; Nakajima, T.; Honda, Y.; Kitao, O.; Nakai, H.; Vreven, T.; Montgomery Jr., J. A.; Peralta, J. E.; Ogliaro, F.; Bearpark, M. J.; Heyd, J.; Brothers, E. N.; Kudin, K. N.; Staroverov, V. N.; Kobayashi, R.; Normand, J.; Raghavachari, K.; Rendell, A. P.; Burant, J. C.; Iyengar, S. S.; Tomasi, J.; Cossi, M.; Rega, N.; Millam, N. J.; Klene, M.; Knox, J. E.; Cross, J. B.; Bakken, V.; Adamo, C.; Jaramillo, J.; Gomperts, R.; Stratmann, R. E.; Yazyev, O.; Austin, A. J.; Cammi, R.; Pomelli, C.; Ochterski, J. W.; Martin, R. L.; Morokuma, K.; Zakrzewski, V. G.; Voth, G. A.; Salvador, P.; Dannenberg, J. J.; Dapprich, S.; Daniels, A. D.; Farkas, Ö.; Foresman, J. B.; Ortiz, J. V.; Cioslowski, J.; Fox, D. J. *Gaussian 09*, Gaussian, Inc.: Wallingford, CT, USA, 2009.

67. Chang, X.-P.; Li, C.-X.; Xie, B.-B.; Cui, G., Photoprotection mechanism of p-methoxy methylcinnamate: a CASPT2 study. *J. Phys. Chem. A* **2015,** *119* (47), 11488-11497.

68. Pedregosa, F.; Varoquaux, G.; Gramfort, A.; Michel, V.; Thirion, B.; Grisel, O.; Blondel, M.; Prettenhofer, P.; Weiss, R.; Dubourg, V., Scikit-learn: Machine learning in Python. *J. Mach. Learn. Res.* **2011,** *12* (Oct), 2825-2830.

69. Geman, S.; Bienenstock, E.; Doursat, R., Neural networks and the bias/variance dilemma. *Neural Comput.* **1992,** *4* (1), 1-58.

70. Naftaly, U.; Intrator, N.; Horn, D., Optimal ensemble averaging of neural networks. *Network-Comp Neural* **1997,** *8* (3), 283-296.

71. Clemen, R. T., Combining forecasts: A review and annotated bibliography. *International journal of forecasting* **1989,** *5* (4), 559-583.

72. Higashi, M.; Truhlar, D. G., Electrostatically Embedded Multiconfiguration Molecular Mechanics Based on the Combined Density Functional and Molecular Mechanical Method. *J. Chem. Theory Comput.* **2008,** *4* (5), 790-803.

73. Collins, M. A., Molecular potential-energy surfaces for chemical reaction dynamics. *Theor. Chem. Acc.* **2002,** *108* (6), 313-324.

74. Hashem, S., Optimal linear combinations of neural networks. *Neural Netw.* **1997,** *10* (4), 599-614.

75. Liu, Y.; Yao, X., Ensemble learning via negative correlation. *Neural Netw.* **1999,** *12* (10), 1399-1404.

76. Voter, A. F.; Montalenti, F.; Germann, T. C., Extending the time scale in atomistic simulation of materials. *Annu. Rev. Mater. Res.* **2002,** *32* (1), 321-346.

77. Shirts, M. R.; Pande, V. S., Mathematical analysis of coupled parallel simulations. *Phys. Rev. Lett.* **2001,** *86* (22), 4983.

78. Joshi, K. L.; Raman, S.; van Duin, A. C., Connectivity-based parallel replica dynamics for chemically reactive systems: from femtoseconds to microseconds. *J. Phys. Chem. Lett.* **2013,** *4* (21), 3792-3797.

79. Perez, D.; Uberuaga, B. P.; Voter, A. F., The parallel replica dynamics method – Coming of age. *Comput. Mater. Sci.* **2015,** *100, Part B*, 90-103.

80. LeCun, Y.; Bengio, Y.; Hinton, G., Deep learning. *Nature* **2015,** *521* (7553), 436-444.

81. Kasha, M., Characterization of electronic transitions in complex molecules. *Discuss. Faraday Soc.* **1950,** *9* (0), 14-19.